\documentclass[12pt, draftclsnofoot, onecolumn]{IEEEtran}

\usepackage{amssymb}
\usepackage{amsmath}
\usepackage{cite}
\usepackage{url}
\usepackage{xcolor}
\usepackage{cite,graphicx,amsmath,amssymb}
\usepackage{subfigure}
\usepackage{fancyhdr}
\usepackage{mdwmath}
\usepackage{mdwtab}
\usepackage{caption}
\usepackage{amsthm}
\usepackage{setspace}
\usepackage{hyperref}
\usepackage{algorithm}
\usepackage{algorithmic}
\hypersetup{colorlinks=false}

\newtheorem{theorem}{Theorem}

\newtheorem{lemma}{Lemma}

\newtheorem{corollary}{Corollary}

\title{NOMA Inspired Interference Cancellation for Integrated Sensing and Communication}

\author{
    \IEEEauthorblockN{Zhaolin Wang\IEEEauthorrefmark{1}, Yuanwei Liu\IEEEauthorrefmark{1}, Xidong Mu\IEEEauthorrefmark{2}, and Zhiguo Ding\IEEEauthorrefmark{3} \\}
    \IEEEauthorblockA{\IEEEauthorrefmark{1}Queen Mary University of London, London, UK. \\ \IEEEauthorrefmark{2}Beijing University of Posts and Telecommunications, Beijing, China. \\\IEEEauthorrefmark{3}The University of Manchester, Manchester, UK \\E-mail: \{zhaolin.wang, yuanwei.liu\}@qmul.ac.uk, muxidong@bupt.edu.cn, zhiguo.ding@manchester.ac.uk}
    }

\begin{document}

\maketitle
\vspace{-2cm}

\begin{abstract}
    A non-orthogonal multiple access (NOMA) inspired integrated sensing and communication (ISAC) system is investigated. A dual-functional 
    base station (BS) serves multiple communication users while sensing multiple targets, by transmitting the \textit{non-orthogonal} superposition of the communication and sensing signals.
    A NOMA inspired interference cancellation scheme is proposed, where part of the dedicated sensing signal is treated as the virtual communication signals 
    to be mitigated at each communication user via successive interference cancellation (SIC).
    Based on this framework, the transmitted communication and sensing signals are jointly optimized to match the desired sensing 
    beampattern, while satisfying the minimum rate requirement and the SIC condition at the communication users. Then, the formulated non-convex optimization problem 
    is solved by invoking the successive convex approximation (SCA) to obtain a near-optimal solution. 
    The numerical results show the proposed NOMA-inspired ISAC system can achieve better performance than the conventional ISAC system and comparable performance 
    to the ideal ISAC system where all sensing interference is assumed to be removed unconditionally.
\end{abstract}


\section{Introduction}
In the past five generations of wireless networks, there is only one envisaged function, i.e., wireless communication. However, 
the beyond fifth generation (B5G) and sixth generation (6G) wireless techniques are regarded as a key enabler for a host of emerging 
applications like unmanned aerial vehicles (UAVs), autonomous driving, virtual and augmented reality, Internet of Vehicles, and smart city,
leading to a trend towards the convergence of multiple functions including communications, sensing, control, etc \cite{letaief2019roadmap, saad2019vision}. Toward this trend, integrated sensing 
and communication (ISAC) is regarded as a promising technique in the B5G and 6G wireless networks and has received heated discussion recently \cite{liu2018toward, liu2018mu, liu2020beamforming, hua2021optimal, liu2021integrated}. 
The goal of ISAC is to merge the two functions in a single system and carry them out jointly and simultaneously, which can provide multiple benefits.
On the one hand, through the dual usage of the hardware facilities and signals for both communication and sensing, it can enhance the spectral and energy efficiency and 
reduce the cost. On the other hand, thanks to such integration, the system can be further benefited from the mutual assistant, such as 
communication-aided sensing and sensing-aided communication.

Given that the ISAC can fully integrate the functions of communication and sensing by sharing the same hardware facilities, signals,
and spectrums, there have been extensive works to jointly design the transceiver towards the objectives of both functions.
The investigation of ISAC starts from the single-antenna technique.
For example, the potential of the orthogonal frequency division multiplexing (OFDM) for simultaneous communication and sensing was studied in \cite{sturm2009ofdm} and \cite{sturm2011waveform}.
Furthermore, given the successful application of the multi-input multi-output (MIMO) technique in both communication \cite{tse2005fundamentals} and radar sensing \cite{li2007mimo}, recent works \cite{liu2018toward, liu2018mu, liu2020beamforming, hua2021optimal} 
begin focusing on its application in ISAC, where multiple beams is generated by exploiting spatial degrees of freedom (DoFs) to serve multiple users and sense multiple targets.
Specifically, the authors in \cite{liu2018toward} derived the closed-form optimal transmission waveforms for minimizing the multiuser interference under 
different radar sensing criteria, while the authors in \cite{liu2018mu} investigated the problem of matching a desired sensing beampattern under the requirements of the communication signal-to-interference-plus-noise ratio (SINR).
However, merely communication signal is employed in both \cite{liu2018toward} and \cite{liu2018mu}, 
which may lead to the beampattern distortion due to the lack of transmit DoFs especially when there are fewer communication users than transmit antennas \cite{liu2020beamforming}.
To address this problem, the authors in \cite{liu2020beamforming} 
introduced a dedicated sensing signal and jointly optimize it with the communication signals to achieve full DoFs for radar sensing.
In \cite{hua2021optimal}, the ability of the dedicated sensing signal was further studied, which theoretically proved and demonstrated that the ISAC system can be significantly benefited from removing 
the sensing interference at the receiver. 

However, the practical scheme for removing the sensing interference was not given in \cite{hua2021optimal}. Note that the superposition of the communication and sensing signals in \cite{liu2020beamforming} and \cite{hua2021optimal}
is a kind of \textit{non-orthogonal} resource sharing, which shares a similar idea with the non-orthogonal multiple access (NOMA) \cite{liu2017non, liu2018multiple}.
For NOMA, successive interference cancellation (SIC) is employed at receivers for partially or totally mitigating the co-channel interference, thus enhancing the performance. This motivates us to develop a NOMA inspired interference cancellation scheme for ISAC.

Against the above background, in this paper, we design a multi-antenna ISAC system carrying out multi-user communication and multi-target sensing, where a dual-functional BS transmit non-orthogonally superimposed 
communication and sensing signals. 
We propose a NOMA inspired interference cancellation scheme by treating part of the sensing signal as the virtual communication signal, which can be mitigated via SIC at each communication user.
Then, we jointly optimize the communication and sensing signals at the BS to match a desired sensing beampattern, subject to the minimum rate requirement and SIC condition at the communication users.
To address this problem, we propose a successive convex approximation (SCA) \cite{dinh2010local} based iterative algorithm.
The numerical results show that the proposed SCA-based algorithm can achieve a performance comparable to the upper limit of performance obtained by semidefinite relaxation (SDR) \cite{luo2010semidefinite}.
It also reveals that the proposed NOMA inspired ISAC system outperforms the conventional ISAC system without capability of sensing interference cancellation 
and has similar performance to the ISAC system with ideal sensing interference cancellation. 

\section{System Model} \label{sec:system_model}

\begin{figure} [t!]
    \centering
    \includegraphics[width=0.8\textwidth]{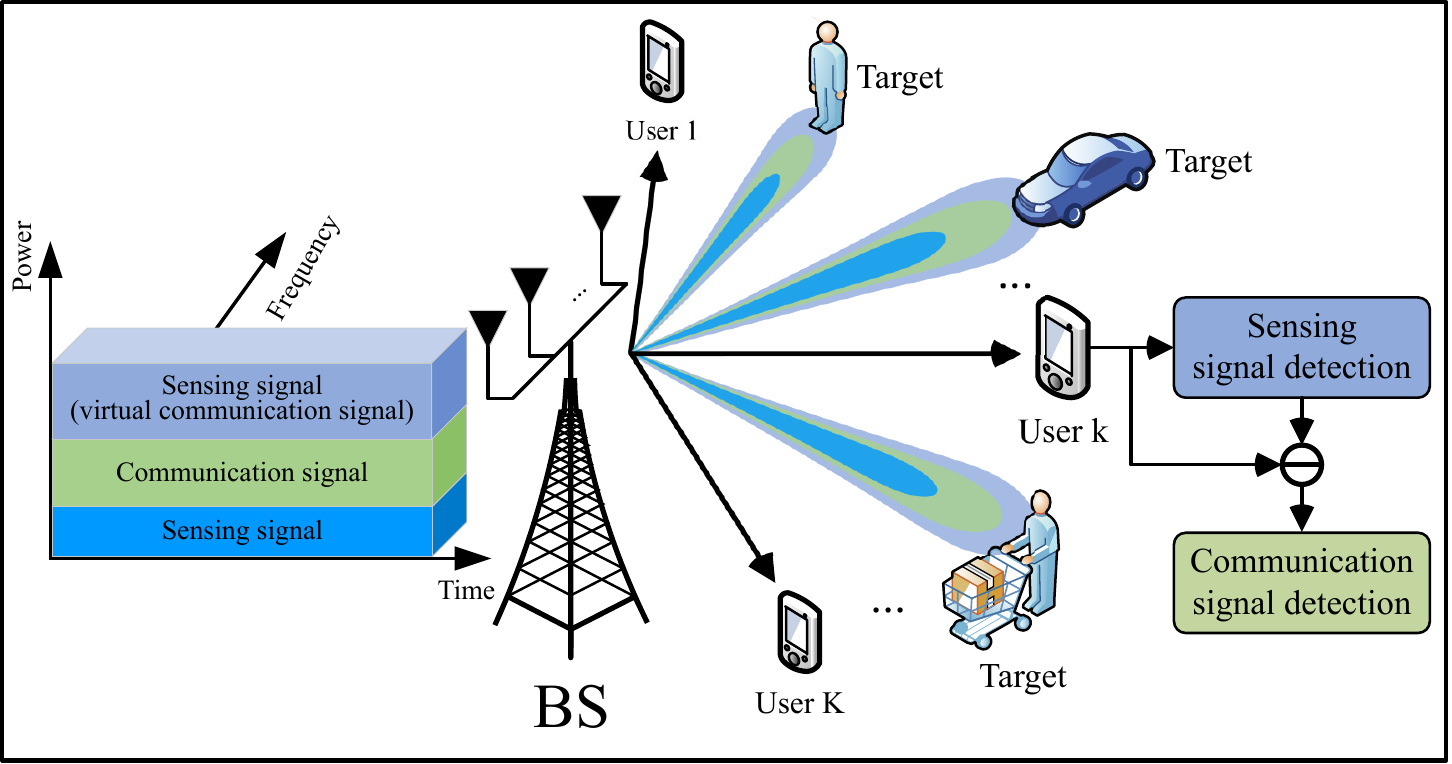}
     \caption{Illustration of the proposed NOMA inspired ISAC system.}
     \label{fig:system_model}
     \vspace{-0.4cm}
 \end{figure}

As shown in Fig. \ref{fig:system_model}, we consider a multi-antenna ISAC system, which consists of a dual-functional $N$-antenna BS, $K$ single-anetnna communication users indexed by $\mathcal{K} = \{1,\dots,K\}$, 
and multiple sensing targets. Similar to the case of \cite{liu2020beamforming} and \cite{hua2021optimal}, the BS transmits the communication signals $\mathbf{w}_i s_i$ for $\forall i \in \mathcal{K}$ and a dedicated sensing signal $\mathbf{r} \in \mathbb{C}^{N \times 1}$, where $\mathbf{w}_i \in \mathbb{C}^{N \times 1}$ denotes the beamformer for 
transmitting the information symbol $s_i \in \mathbb{C}$ to user $i$. The information symbols $\{s_i\}_{i \in \mathcal{K}}$ are assumed to be independent as well as with zero mean and unit power. The dedicated sensing signal is assumed to be independent with the information symbols $\{s_i\}_{i \in \mathcal{K}}$ and with 
the covariance matrix $\mathbf{R}_{\mathbf{r}} = \mathbb{E}[\mathbf{r}\mathbf{r}^H] \succeq 0$. Thus, the transmitted signal at the BS is given by     
\begin{equation}
    \mathbf{x} = \sum_{i \in \mathcal{K}} \mathbf{w}_i s_i + \mathbf{r}.
\end{equation}
The dedicated sensing signal is assumed to exploit multi-beam transmission, i.e., the covariance matrix $\mathbf{R}_{\mathbf{r}}$ 
is with general rank. 
By exploiting the eigenvalue decomposition of $\mathbf{R}_{\mathbf{r}}$, the dedicated sensing signal can be decomposed into multiple 
sensing beams, i.e.,
\begin{equation}
    \mathbf{R}_{\mathbf{r}} = \sum_{i=1}^{\mathrm{rank}(\mathbf{R}_{\mathbf{r}})} \lambda_i \mathbf{v}_{i} \mathbf{v}_{i}^H  = \sum_{i=1}^{\mathrm{rank}(\mathbf{R}_{\mathbf{r}})} \mathbf{w}_{r,i} \mathbf{w}_{r,i}^H
\end{equation}
where $\lambda_i \in \mathbb{R}$ is the eigenvalue and $\mathbf{v}_i \in \mathbb{C}^{N \times 1}$ is the corresponding eigenvector. The vector $\mathbf{w}_{r,i} = \sqrt{\lambda_i} \mathbf{v}_i$
is the transmission beamformer for the dedicated sensing signal.
Inspired by NOMA, we embed information into part of the dedicated sensing signal and treat the information-embedded sensing signal as the virtual communication signals on the top of the real communication signals.
Thus, the sensing signal can be rewritten as 
\begin{equation}
    \mathbf{r} = \sum_{i \in \mathcal{M}} \mathbf{w}_{r,i} r_i + \tilde{\mathbf{r}},
\end{equation}
where $\mathcal{M} = \{ 1,\dots,M \}, 1 \le M < \mathrm{rank}(\mathbf{R}_{\mathbf{r}})$ and the sensing symbols $\{r_i\}_{i \in \mathcal{M}}$ are independent as well as with zero mean and unit power. 
We also assume that $\{r_i\}_{i \in \mathcal{M}}$ are independent with $ \tilde{\mathbf{r}}$. 
Thus, the covariance matrix $\mathbf{R}_{\tilde{\mathbf{r}}}$ of the rest of sensing signal $\tilde{\mathbf{r}}$ given by 
\begin{align}
    \mathbf{R}_{\tilde{\mathbf{r}}} = \mathbb{E}\Big[ \big(\mathbf{r} - \sum_{i \in \mathcal{M}} \mathbf{w}_{r,i} r_i\big) \big(\mathbf{r} - \sum_{i \in \mathcal{M}} \mathbf{w}_{r,i} r_i\big)^H \Big] =  \sum_{i=M+1}^{\mathrm{rank}(\mathbf{R}_{\mathbf{r}})} \mathbf{w}_{r,i} \mathbf{w}_{r,i}^H,
\end{align}
which indicates that the matrix $\mathbf{R}_{\tilde{\mathbf{r}}}$ is also with general rank.

\subsubsection{Communication Model}
Given the transmitted signal, the received signal at user $k$ is given by
\begin{align}
    y_k = \mathbf{h}_k^H \mathbf{w}_k s_k + \underbrace{\mathbf{h}_k^H \sum_{i \in \mathcal{K}, i \neq k} \mathbf{w}_i s_i}_{\text{inter-user interference}} + \underbrace{\mathbf{h}_k^H \sum_{i \in \mathcal{M}} \mathbf{w}_{r,i} r_i}_{\scriptstyle \text{sensing interference} \atop \scriptstyle \text{removed by SIC}} + \underbrace{\mathbf{h}_k^H \tilde{\mathbf{r}}}_{\scriptstyle \text{sensing interference} \atop \scriptstyle \text{not removed by SIC}} + n_k.
\end{align} 
where $\mathbf{h}_k \in \mathbb{C}^{N \times 1}, \forall k \in \mathcal{K}$ denotes the BS-user channel and $n_k \in \mathbb{C}, \forall k \in \mathcal{K}$ denotes the circularly symmetric
complex Gaussian noise with variance $\sigma_n^2$. 
By exploiting SIC, the virtual communication signals can be eliminated at the communication user.
Thus, the achievable rate at user $k$ is given by
\begin{equation}
    R_k = \log_2 \left( 1 + \frac{| \mathbf{h}_k^H \mathbf{w}_k|^2}{ \sum_{i \in \mathcal{K}, i \neq k} |\mathbf{h}_k^H \mathbf{w}_i|^2 +  \mathbf{h}_k^H \mathbf{R}_{\tilde{\mathbf{r}}} \mathbf{h}_k + \sigma_n^2} \right).
\end{equation}
The following condition need to be satisfied to carry out SIC at user $k$: 
\begin{equation} \label{eqn:constraint_SIC}
    |\mathbf{h}_k^H \mathbf{w}_{r,i}|^2 \ge |\mathbf{h}_k^H \mathbf{w}_k|^2, \forall i \in \mathcal{M},
\end{equation}
which means the interference from the virtual communication signal can be eliminated only if its power is higher than that of the desired communication signal. 

\subsubsection{Sensing Model}
For sensing function, the key metric is the transmit beampattern, which is determined by the transmit covariance matrix \cite{stoica2007probing}. The covariance matrix 
is given by
\begin{equation}
    \mathbf{R} = \mathbb{E}[\mathbf{x}\mathbf{x}^H] = \sum_{i \in \mathcal{K}} \mathbf{w}_i \mathbf{w}_i^H + \sum_{i \in \mathcal{M}} \mathbf{w}_{r,i} \mathbf{w}_{r,i}^H + \mathbf{R}_{\tilde{\mathbf{r}}}.
\end{equation}
Then, the beampattern of transmitted signal in the direction $\theta_l$ is given by 
\begin{equation}
    P(\theta_l) = \mathbf{a}^H(\theta_l) \mathbf{R} \mathbf{a}(\theta_l),
\end{equation} 
where $ \mathbf{a}(\theta_l) = [1, e^{j\frac{2\pi}{\lambda}d\sin({\theta_l})},...,e^{j\frac{2\pi}{\lambda}d(N-1)\sin({\theta_l})}]^T$ denotes the steering vector when the uniform linear array (ULA) is used as the antenna employment at the BS.
In practice, the desired sensing beampattern is designed according to the sensing requirements \cite{stoica2007probing}. For example, if the sensing system
has no information about target and works in the detecting mode, an isotropic beampattern is desired, i.e., the power is uniformly distributed among all directions.
However, when the sensing system has prior information of targets and works in the tracking mode, the beampattern is expected to have the dominant peaks in the target directions.


\section{Problem Formulation and Proposed Solution} \label{sec:solution}

\subsection{Problem Formulation}
Let $\{\phi(\theta_l)\}_{l=1}^L$ denotes the desired sensing beampattern over an angular grid $\{\theta_l\}_{l=1}^L$ covering the detector's angular range $[-\pi/2, \pi/2]$.
We aim to jointly optimize the communication signal and the dedicated sensing signal such that matching error 
between the transmit beampattern and the desired sensing beampattern is minimized in the least squares sense, subject to the constraints of the minimum communication rate $R_{\min,k}, \forall k \in \mathcal{K}$, the SIC conditions, and the total transmit power budget.
The resultant optimization problem is formulated as 
\vspace{-0.4cm}
\begin{subequations} \label{problem:dl}
    \begin{align}
      \min_{\scriptstyle \delta, \{\mathbf{w}_i\}_{i \in \mathcal{K}}, \atop \scriptstyle \{\mathbf{w}_{r,i}\}_{i \in \mathcal{M}}, \mathbf{R}_{\tilde{\mathbf{r}}} } & \sum_{l=1}^{L}\left| \delta \phi(\theta_l) - \mathbf{a}^H(\theta_l) \mathbf{R} \mathbf{a}(\theta_l)  \right|^2 \\
      \label{constraint:min_rate}
      \mathrm{s.t.} \quad & R_k \ge R_{\min,k}, \forall k \in \mathcal{K},\\
      \label{constraint:SIC}
      & |\mathbf{h}_k^H \mathbf{w}_{r,i}|^2 \ge |\mathbf{h}_k^H \mathbf{w}_k|^2, \forall i \in \mathcal{M}, \forall k \in \mathcal{K} \\
      \label{constraint:uniform_power}
      & \mathrm{Tr}(\mathbf{R}) \le P_t, \\
      & \mathbf{R}_{\tilde{\mathbf{r}}} \succeq 0,
    \end{align}
\end{subequations}

\noindent where $\delta$ denotes the scaling factor and $P_t$ denotes the total transmit power budget. The main challenges for solving this problem is summarized as follows: firstly, the non-concave form of 
the achievable rate makes the constraint \eqref{constraint:min_rate} non-convex; secondly, the quadratic form of the beamformers makes the constraints
\eqref{constraint:SIC} non-convex.

\subsection{Proposed Solution}
In this section, we proposed an iterative algorithm based on SCA to solve problem \eqref{problem:dl}. We firstly define the 
auxiliary varaibles $\mathbf{W}_k = \mathbf{w}_k \mathbf{w}_k^H, \forall k \in \mathcal{K}$, where $\mathbf{W}_i \succeq 0$ and $\mathrm{rank}(\mathbf{W}_i)=1$,   
and $\mathbf{W}_{r,i} = \mathbf{w}_{r,i} \mathbf{w}_{r,i}^H, \forall i \in \mathcal{M}$, where $\mathbf{W}_{r,i} \succeq 0$ and $\mathrm{rank}(\mathbf{W}_{r,i})=1$.
Then, problem \eqref{problem:dl} can be reformulated as
\vspace{-0.4cm}
\begin{subequations} \label{problem:dl_SDR}
    \begin{align}
      \min_{\scriptstyle \delta, \{\mathbf{W}_i\}_{i \in \mathcal{K}}, \atop \scriptstyle \{\mathbf{W}_{r,i}\}_{i \in \mathcal{M}}, \mathbf{R}_{\tilde{\mathbf{r}}} } & \sum_{l=1}^{L}\left| \delta \phi(\theta_l) - \mathbf{a}^H(\theta_l) \mathbf{R} \mathbf{a}(\theta_l)  \right|^2 \\
      \label{constraint:min_rate_2}
      \mathrm{s.t.} \quad &  \mathbf{h}_k^H \mathbf{W}_k \mathbf{h}_k - \Gamma_k N_k \ge 0, \forall k \in \mathcal{K},\\
      \label{constraint:SIC_2}
      & \mathbf{h}_k^H \mathbf{W}_{r,i} \mathbf{h}_k \ge \mathbf{h}_k^H \mathbf{W}_k \mathbf{h}_k, \forall i \in \mathcal{M}, \forall k \in \mathcal{K} \\
      \label{constraint:uniform_power_2}
      & \mathrm{Tr}(\mathbf{R}) \le P_t, \\
      \label{constraint:rank_comm}
      & \mathbf{W}_k \succeq 0, \mathrm{rank}(\mathbf{W}_k) = 1, \forall k \in \mathcal{K}\\
      \label{constraint:rank_sens}
      & \mathbf{W}_{r,i} \succeq 0, \mathrm{rank}(\mathbf{W}_{r,i}) = 1, \forall i \in \mathcal{M}, \\
      \label{constraint:radar_covariance}
      & \mathbf{R}_{\tilde{\mathbf{r}}} \succeq 0,
    \end{align}
\end{subequations}
where $\Gamma_k = 2^{R_{\min,k}}-1$ and $N_k = \sum_{i \in \mathcal{K}, i \neq k} \mathbf{h}_k^H \mathbf{W}_i \mathbf{h}_k + \mathbf{h}_k^H \mathbf{R}_{\tilde{\mathbf{r}}} \mathbf{h}_k + \sigma_n^2$. This problem is still non-convex only due to the rank-one constraints \eqref{constraint:rank_comm} and \eqref{constraint:rank_sens}.
By omitting the rank-one constraints, we obtain the SDR of this problem, which is a convex semidifinite programe (SDP). Thus, the global optimum to the SDR 
can be efficiently obtained through convex solvers like CVX \cite{cvx}. Nevertheless, the global optimum of the SDR may not be the global optimum of problem \eqref{problem:dl_SDR}, 
since the matrices $\mathbf{W}_k, \forall k \in \mathcal{K}$ and $\mathbf{W}_{r,i}, \forall i \in \mathcal{M}$ are not guaranteed to be rank-one. 
Generally,  given the general-rank solution obtained by SDR, the eigenvalue decomposition or the Gaussian randomization need to be exploited to reconstruct a rank-one solution \cite{luo2010semidefinite}. 
However, the feasibility of the reconstructed rank-one solution is not guaranteed. To avoid this drawback, we employ a penalty-based scheme to 
transform the rank-one constraints to a penalty term in the objective function \cite{mu2021noma} and solve it by invoking SCA \cite{dinh2010local}.
\vspace{-0.2cm}

Firstly, the rank-one constraints are equivalent to 
\vspace{-0.2cm}
\begin{subequations}
  \begin{align}
    \| \mathbf{W}_k \|_* - \| \mathbf{W}_k \|_2 &= 0, \forall k \in \mathcal{K}, \\
    \| \mathbf{W}_{r,i} \|_* - \| \mathbf{W}_{r,i} \|_2 &= 0, \forall i \in \mathcal{M},
  \end{align}
\end{subequations}
where $\| \cdot \|_*$ and $\| \cdot \|_2$ denote the nuclear norm and spectral norm, respectively. Since the matrices $\mathbf{W}_k, \forall k \in \mathcal{K}$ and 
$\mathbf{W}_{r,i}, i \in \mathcal{M}$ are positive semidifinite, 
the inequalities $\| \mathbf{W}_k \|_* - \| \mathbf{W}_k \|_2 > 0$ and  
$\| \mathbf{W}_{r,i} \|_* - \| \mathbf{W}_{r,i} \|_2 > 0$ must hold if the matrices are not rank-one. 
Thus, by minimizing the difference 
between the nuclear norm and spectral norm, we can obtain the metrics that are nearly rank-one. It can be achieved by introducing a
penalty term to the objective function of problem \eqref{problem:dl_SDR}, yielding the following optimization problem:
\vspace{-0.3cm}
\begin{subequations} \label{problem:penalty_non_convex}
  \begin{align}
    \min_{\scriptstyle \delta, \{\mathbf{W}_i\}_{i \in \mathcal{K}}, \atop \scriptstyle \{\mathbf{W}_{r,i}\}_{i \in \mathcal{M}}, \mathbf{R}_{\tilde{\mathbf{r}}} } & \sum_{l=1}^{L}\left| \delta \phi(\theta_l) - \mathbf{a}^H(\theta_l) \mathbf{R} \mathbf{a}(\theta_l)  \right|^2 \nonumber\\
    &+ \frac{1}{\rho} \sum_{k \in \mathcal{K}} \big(\| \mathbf{W}_k \|_* - \| \mathbf{W}_k \|_2 \big) + \frac{1}{\rho} \sum_{i \in \mathcal{M}} \big(\| \mathbf{W}_{r,i} \|_* - \| \mathbf{W}_{r,i} \|_2 \big) \\
    \label{semidefinite_comm}
    \mathrm{s.t.} \quad & \mathbf{W}_k \succeq 0, \forall k \in \mathcal{K}, \mathbf{W}_{r,i} \succeq 0, \forall i \in \mathcal{M}, \\
    \label{semidefinite_sense_2}
    & \mathbf{R}_{\tilde{\mathbf{r}}} \succeq 0, \\
    & \eqref{constraint:min_rate_2} - \eqref{constraint:uniform_power_2},
  \end{align}
\end{subequations}

\noindent where $\rho$ is the regularization parameter. If $\rho=0$, i.e., $1/\rho \rightarrow +\infty$, the solution $\mathbf{W}_k, \forall k \in \mathcal{K}$ 
and $\mathbf{W}_{r,i}, \forall i \in \mathcal{M}$ will be exactly rank-one. However, in this case, the beampattern matching error cannot be 
sufficiently minimized as the objective function is dominated by the penalty term. To address this problem, we can initialize $\rho$ with a large value to 
find a good start point with respect to the beampattern matching error. Then, by gradually reducing $\rho$ to a sufficiently small 
value, the solution $\mathbf{W}_k, \forall k \in \mathcal{K}$ and $\mathbf{W}_{r,i}, \forall i \in \mathcal{M}$ will gradually approach to be rank-one.
Toward this idea, the parameter $\rho$ is updated following
\vspace{-0.4cm}
\begin{equation}
  \rho = \varepsilon \rho, 0 < \varepsilon < 1.
\end{equation} 
This procedure is terminated when the penalty term is smaller than a pre-defined threshold $\epsilon_1 \ge 0$, i.e.,
\begin{equation}
   \sum_{k \in \mathcal{K}} \big(\| \mathbf{W}_k \|_* - \| \mathbf{W}_k \|_2 \big) + \sum_{i \in \mathcal{M}} \big(\| \mathbf{W}_{r,i} \|_* - \| \mathbf{W}_{r,i} \|_2 \big) \le \epsilon_1.
\end{equation} 
\vspace{-0.6cm}



However, problem \eqref{problem:penalty_non_convex} is still non-convex due to the terms $-\| \mathbf{W}_k \|_2 $ and $-\| \mathbf{W}_{r,i} \|_2 $, 
which can be solved by invoking SCA. A convex upper bound of the non-convex term $-\| \mathbf{W}_k \|_2 $ can be obtained via the first-order Taylor expansion at the point $\mathbf{W}_k^n$, which is given by
\begin{align} 
   \tilde{\mathbf{W}}_k^n \triangleq -\| \mathbf{W}_k^n \|_2 - \mathrm{Tr} \big[ \mathbf{u}_{\max,k}^n (\mathbf{u}_{\max,k}^n)^H \left( \mathbf{W}_k - \mathbf{W}_k^n \right) \big], 
\end{align}
where $\mathbf{u}_{\max,k}^n$ denotes the eigenvector corresponding to the largest eigenvalue of the matrix $\mathbf{W}_k^n$. The upper bounds $\tilde{\mathbf{W}}_{r,i}^n$ 
of the term $-\|\mathbf{W}_{r,i}\|_2$ can be defined similarly as $\tilde{\mathbf{W}}_k^n$. Thus, problem \eqref{problem:penalty_non_convex}
can be reformulated as
\begin{subequations} \label{problem:penalty_convex}
  \begin{align}
    \min_{\scriptstyle \delta, \{\mathbf{W}_i\}_{i \in \mathcal{K}}, \atop \scriptstyle \{\mathbf{W}_{r,i}\}_{i \in \mathcal{M}}, \mathbf{R}_{\tilde{\mathbf{r}}} } & \sum_{l=1}^{L}\left| \delta \phi(\theta_l) - \mathbf{a}^H(\theta_l) \mathbf{R} \mathbf{a}(\theta_l)  \right|^2 \nonumber\\
    &+ \frac{1}{\rho} \sum_{k \in \mathcal{K}} \big(\| \mathbf{W}_k \|_* + \tilde{\mathbf{W}}_k^n \big) + \frac{1}{\rho} \sum_{i \in \mathcal{M}} \big(\| \mathbf{W}_{r,i} \|_* + \tilde{\mathbf{W}}_{r,i}^n \big) \\
    \mathrm{s.t.} \quad & \eqref{constraint:min_rate_2} - \eqref{constraint:uniform_power_2}, \eqref{semidefinite_comm}, \eqref{semidefinite_sense_2},
  \end{align}
\end{subequations}
which is a quadratic semidefinite program (QSDP) and can be efficiently solved by the convex solvers like CVX \cite{cvx}. Based on the procedures above, the overall algorithm 
for solve prblem \eqref{problem:dl} is summarized in \textbf{Algorithm \ref{alg:A}}. Given the solution accurancy $\epsilon$, the computational burden of this algorithm mainly arise from solving 
the SDP \eqref{problem:penalty_convex} with complexity $\mathcal{O}((K+M+1)^{6.5} N^{6.5}\log(1/\epsilon))$ \cite{liu2020beamforming} and the number of inner and outer iterations, which are $I_i$ and $I_o$ respectively. Thus, 
the complexity of \textbf{Algorithm \ref{alg:A}} is $\mathcal{O}(I_o I_i (K+M+1)^{6.5} N^{6.5}\log(1/\epsilon))$. 
\begin{algorithm}[htb]
  \caption{Proposed iterative SCA-based algorithm for solving problem \eqref{problem:dl}.}
  \label{alg:A}
  \begin{algorithmic}[1]
      \REQUIRE{$\rho$, $\mathbf{W}_k^0, \forall k \in \mathcal{K}, \mathbf{W}_{r,i}^0, \forall i \in \mathcal{M}$;}
      \REPEAT
      \STATE{$n \gets 0$};
        \REPEAT
          \STATE{Based on given $\mathbf{W}_k^n$ and $\mathbf{W}_{r,i}^n$, calculate the optimal $\mathbf{W}_k^\star$ and $\mathbf{W}_{r,i}^\star$ by solving problem \eqref{problem:penalty_convex} for $\forall k \in \mathcal{K}$ and $\forall i \in \mathcal{M}$;}
          \STATE{$\mathbf{W}_k^{n+1} \gets \mathbf{W}_k^\star, \forall k \in \mathcal{K}$;}
          \STATE{$\mathbf{W}_{r,i}^{n+1} \gets \mathbf{W}_{r,i}^\star, \forall i \in \mathcal{M}$;}
          \STATE{$n \gets n+1$;}
        \UNTIL{the fractional reduction of the objective function value falls below a predefined threshold $\epsilon_2$; }
      \STATE{$\mathbf{W}_k^0 \gets \mathbf{W}_k^\star, \forall k \in \mathcal{K}$;}
      \STATE{$\mathbf{W}_{r,i}^0 \gets \mathbf{W}_{r,i}^\star, \forall i \in \mathcal{M}$;}
      \STATE{$\rho \gets \varepsilon \rho$;}
      \UNTIL{the value of the penalty term falls below a predefined threshold $\epsilon_1$;}

  \end{algorithmic}
\end{algorithm}

\section{Numerical Results} \label{sec:results}
 In this section, the numerical results are provided for characterizing the proposed NOMA inspired ISAC system. We assume a dual-functional BS 
 equipped with an $N=8$ antenna ULA with half wavelength spacing, serving $K=5$ communication users and sensing $3$ targets in the directions 
 $\Phi = \{-60^\circ, 0^\circ, 60^\circ\}$.
 Given the directions of sensing targets, the desired beampattern is given by 
 \begin{equation}
     \phi (\theta_l) = \begin{cases}
         1, &\theta_l \in [\varphi - \frac{\Delta}{2}, \varphi - \frac{\Delta}{2}], \forall \varphi \in \Phi, \\
         0, &\mathrm{otherwise},
     \end{cases}
 \end{equation}
 where $\Delta$ is the desired beam width, which is set to $10^\circ$ in the simulation.
 We also assume that only one beam of the dedicated sensing signal is embedded with information and eliminated by SIC at the communication users, i.e., $M=1$. 
 Furthermore, the transmit power budegt at BS and the noise power at the communication users are set to $P_t = 20$ dBm and 
 $\sigma_n^2 = -80$ dBm, respectively. The channels between BS and communication users are assumed to experience Rayleigh fading and $80$ dB pathloss.
 The initial penalty factor is set to $\rho = 10^2$ and its reduction factor is set to $\varepsilon = 0.2$. The convergence thresholds of 
 the inner and outer iterations are set to $\epsilon_2 = 10^{-2}$ and $\epsilon_1 = 10^{-4}$, respectively. The minimum rate requirement is
 assumed to be the same at all communication users, i.e., $R_{\min,k} = R_{\min}, \forall k \in \mathcal{K}$.  
 The results are obtained via Monte Carlo simulation over $50$ random channel realizations.

 \subsection{Benchmark Schemes}
 For comparison, we consider  the following four benchmark schemes:
 \begin{itemize}
    \item \textbf{Ideal sensing interference cancellation (Ideal ISAC) \cite{hua2021optimal}}: In this scheme, the receiver at communication users can perfectly remove the sensing interference with no constraint. 
    Therefore, the achievable rate at user $k$ is given by 
    \begin{equation}
        R_k = \log_2 \left( 1 + \frac{| \mathbf{h}_k^H \mathbf{w}_k|^2}{ \sum_{i \in \mathcal{K}, i \neq k} |\mathbf{h}_k^H \mathbf{w}_i|^2 + \sigma_n^2} \right)
    \end{equation} 
    The global optimum of the resultant beampattern matching error minimization problem can be obtained via SDR \cite{hua2021optimal}.
    
    \item \textbf{No sensing interference cancellation (Conventional ISAC) \cite{liu2020beamforming}}: In this scheme, the receiver at communication users has no ability to eliminate the sensing interference.
    Thus, the achievable rate at user $k$ is given by 
    \begin{equation}
        R_k \! = \! \log_2 \! \left( 1 \! + \! \frac{| \mathbf{h}_k^H \mathbf{w}_k|^2}{ \sum_{i \in \mathcal{K}, i \neq k} |\mathbf{h}_k^H \mathbf{w}_i|^2  \! +  \! \mathbf{h}_k^H \mathbf{R}_{\mathbf{r}} \mathbf{h}_k \! + \! \sigma_n^2} \right),
    \end{equation} 
    where $\mathbf{R}_{\mathbf{r}}$ is the covariance matrix of the overall dedicated sensing signal $\mathbf{r}$.
    Similarly, by exploiting SDR, the globally optimum of the resultant optimization problem can be obtained \cite{liu2020beamforming}.

    \item \textbf{Communication signal only \cite{liu2018mu}}: In this scheme, the BS only transmit the communication signals. Thus, the covariance matrix of the transmitted signal is 
    \begin{equation}
        \mathbf{R} = \sum_{i \in \mathcal{K}} \mathbf{w}_i \mathbf{w}_i^H.
    \end{equation}
    The achievable at user $k$ is given by
    \begin{equation}
        R_k = \log_2 \left( 1 + \frac{| \mathbf{h}_k^H \mathbf{w}_k|^2}{ \sum_{i \in \mathcal{K}, i \neq k} |\mathbf{h}_k^H \mathbf{w}_i|^2 + \sigma_n^2} \right)
    \end{equation}
    Then, the resultant optimization problem can be solved by the proposed \textbf{Algorithm \ref{alg:A}}.

    \item \textbf{Proposed SIC scheme via SDR}: In this scheme, we directly solve the SDR of problem \eqref{problem:dl_SDR} 
    and keep the general-rank solution. As the rank-one constraints are relaxed in SDR, the feasibility region is enlarged. 
    Thus, a performance upper bound can be obtained.
 \end{itemize}

 \subsection{Convergence of the Proposed Algorithm}
 \begin{figure} [t!]
    \centering
    \includegraphics[width=0.5\textwidth]{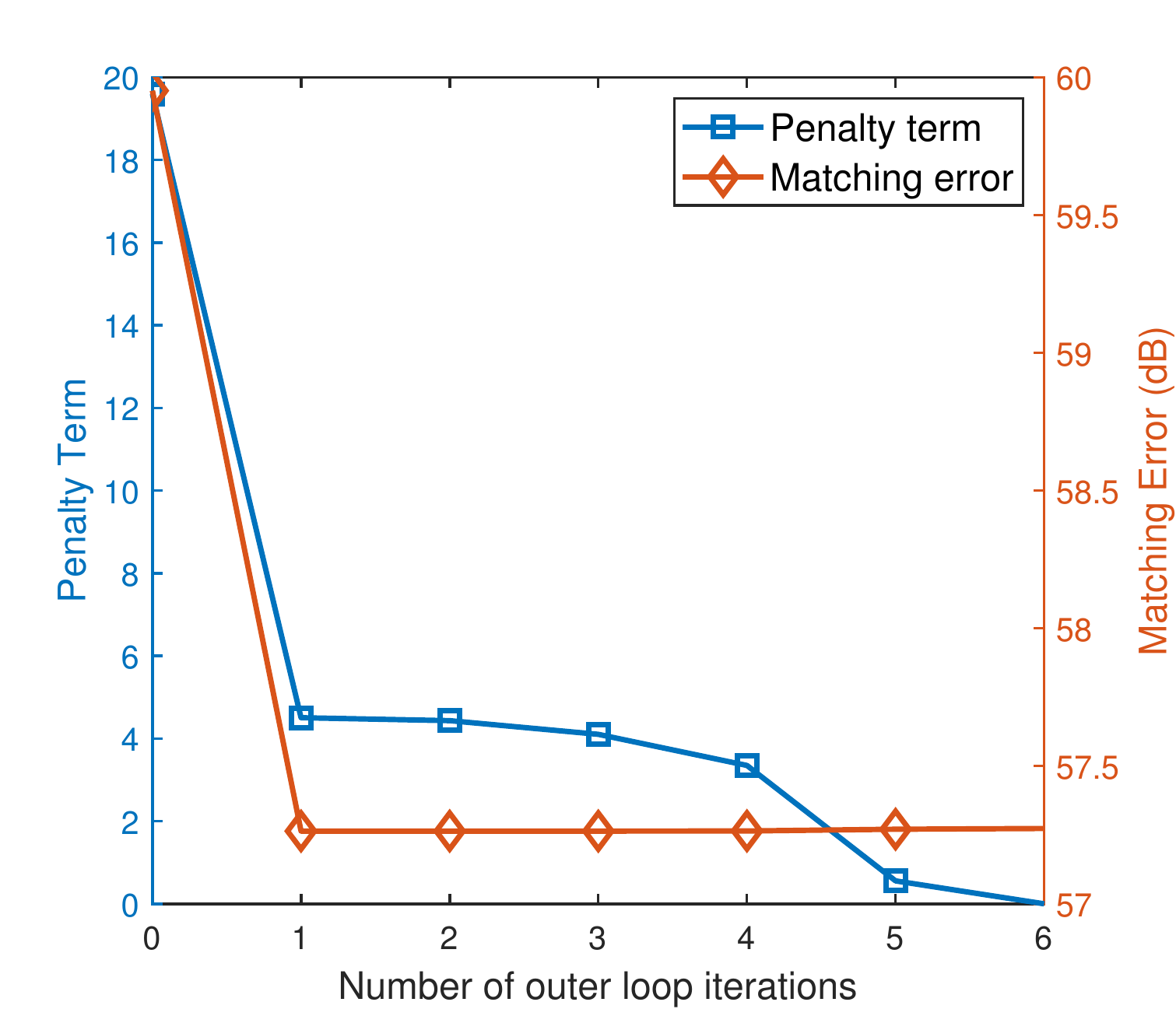}
     \caption{Convergence of the proposed SCA-based algorithm when $R_{\min}=4.5$ bit/s/Hz.}
     \label{fig:convergence}
 \end{figure}
 We demonstrate the convergence behavior of the proposed algorithm over a specific channel realization when $R_{\min}=4.5$ bit/s/Hz in Fig. \ref{fig:convergence}.
 On the one hand, we can observe that the matching error can quickly converge to a stable value, while the penalty term is gradually reduced to almost zero as the number of outer loop iterations increase,
 which reveals that the resultant $\mathbf{W}_{r,i}$ is nearly rank-one. On the other hand, it can also be seen that the matching error is not effected by the reduction of the penalty terms, which indicate the efficiency of the proposed algorithm.

 \subsection{Matching Error versus Minimum Rate}
 \begin{figure} [t!]
    \centering
    \includegraphics[width=0.5\textwidth]{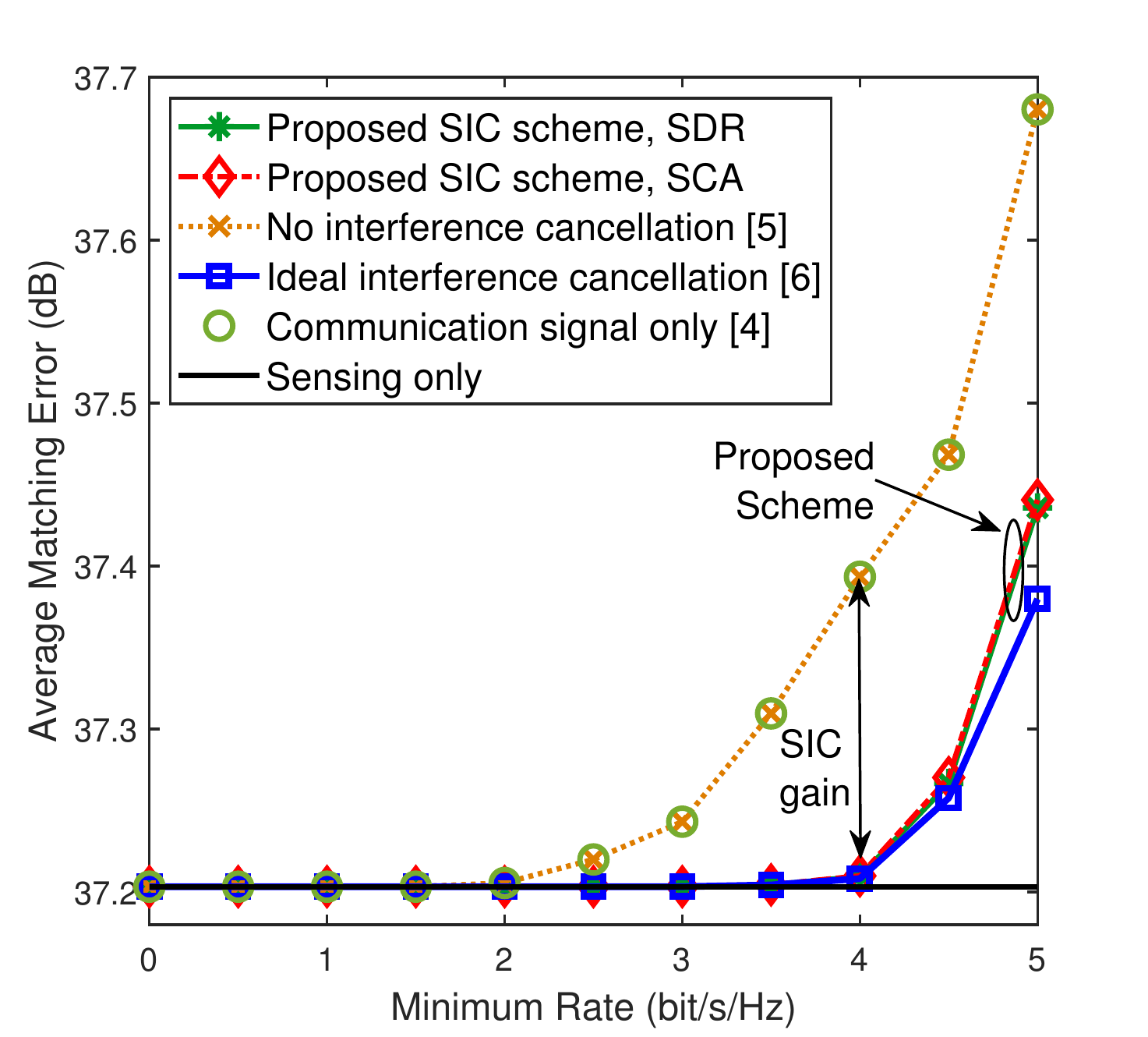}
     \caption{Average matching error versus the minimum rate.}
     \label{fig:error_vs_rate}
 \end{figure}

 In Fig. \ref{fig:error_vs_rate}, we present the average beampattern matching error versus the minimum rate $R_{\min}$.
 It is observed that there is only a negligible gap between the results obtained by the proposed SCA-based algorithm and SDR, 
 which indicates the efficiency of the proposed algorithm. Therefore, although the SCA-based algorithm is 
 sub-optimal, it can provide a good reference of the globally optimal result. It can also be seen that the proposed NOMA inspired ISAC system outperforms 
 the conventional ISAC system without sensing interference cancellation, while the performance of the conventional ISAC system is almost the same as the ISAC system merely transmitting communication signals. 
 This is because compared with conventional ISAC system, where 
 the power of sensing signal is rapidly reduced and even becomes zero when $R_{\min} \ge 2.5$ bit/s/Hz in the simulation, the counterpart in the
 NOMA inspired ISAC system is much larger, which can provide more DoFs for sensing. Moreover, as illustrated in Fig. \ref{fig:error_vs_rate},
 the matching error achieved by the NOMA inspired SIC scheme is close to the that achieved by the ideal interference cancellation scheme, especially when the $R_{\min}$
 is not high. This is because in the simulation, one can observe that the power of the sensing signal is dominated by the largest eigenvalue.
 Thus, only removing $M=1$ sensing beam related to the largest eigenvalue will not lead to the significant performance degradation.
 However, the gap between the NOMA inspired SIC scheme and the ideal interference cancellation scheme becomes larger when $R_{\min}$ is high.
 The reason is that in the ideal ISAC system, the sensing signal only needs to approach the desired sensing beampattern, while in the NOMA inspired ISAC system, 
 it also needs to satisfy the SIC conditions, which hinders the system performance.

 \begin{figure} [t!]
    \centering
    \includegraphics[width=0.5\textwidth]{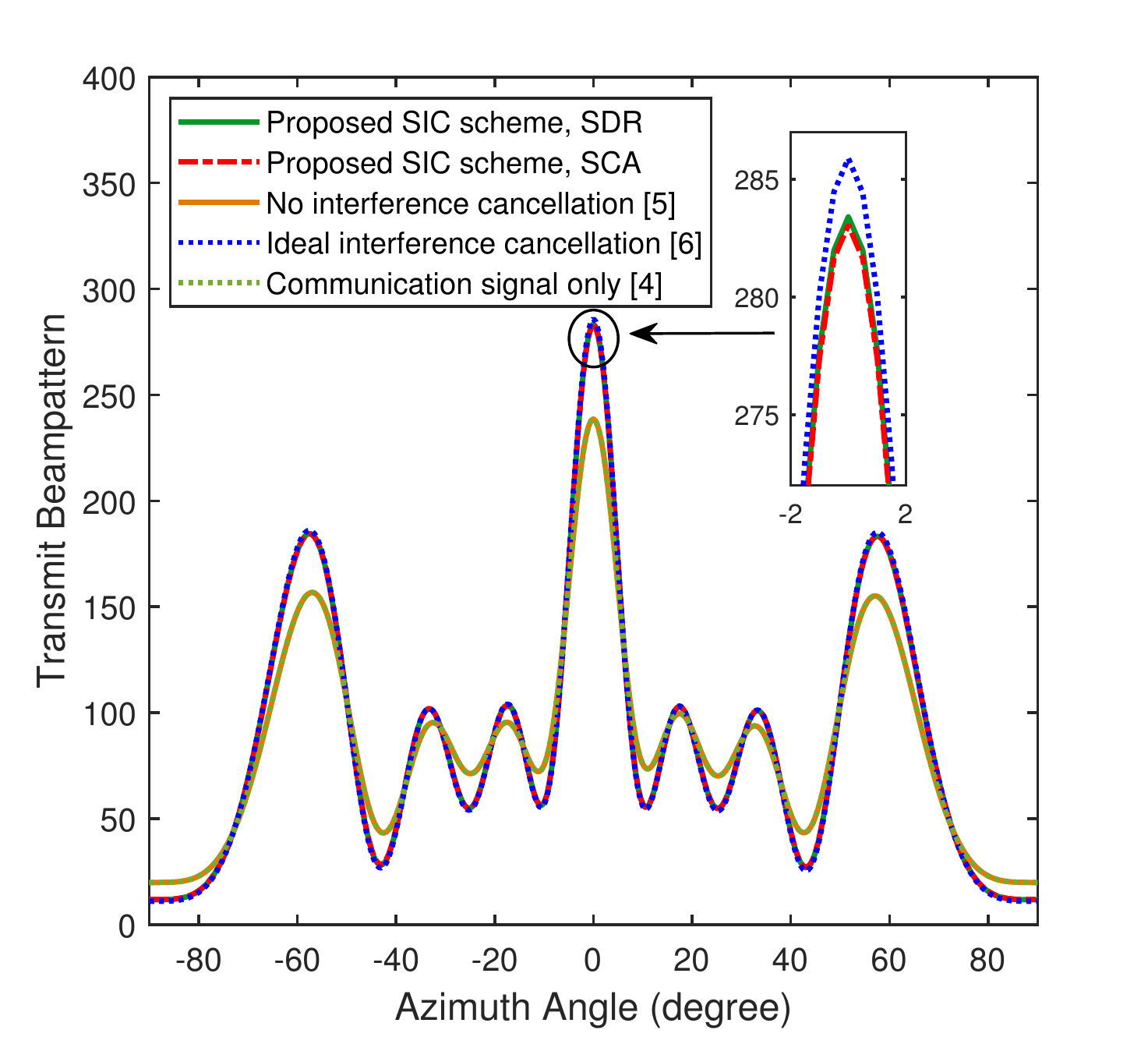}
     \caption{Transmit beampattern with $R_{\min} = 4.5$ bit/s/Hz. }
     \label{fig:beampattern}
 \end{figure}

\subsection{Transmit Beampattern}
In Fig. \ref{fig:beampattern}, we compare the transmit beampatterns obtained by different schemes when $R_{\min} = 4.5$ bit/s/Hz.
Firstly, it can be seen that there is a negligible difference between the beampatterns obtained by the proposed NOMA inspired SIC scheme
via SDR and SCA-based methods. It can also be observed that both proposed scheme and benchmarks can achieve dominant peaks in the target directions.
However, the proposed NOMA inspired ISAC system outperforms the conventional ISAC system, where there is a
noticeable gain of power in the target directions and less power leakage in the undesired directions.
Moreover, there is only a slight gap between the NOMA inspired ISAC system and ideal ISAC system in terms of the beampattern, 
which further verify the efficiency of the proposed scheme. 

\section{Conclusion} \label{sec:conclusion}
In this paper, we proposed a NOMA inspired ISAC system, where the SIC scheme is applied to mitigate the interference to the communication users introduced by
the dedicated sensing signal. In order to support dual-functions, the beamformers for the communication and sensing were jointly designed to 
minimize the beampattern matching error under the constraints of minimum communication rate and the SIC conditions. Given the non-convexity of the resultant optimization problem, 
an iterative algorithm was proposed by invoking SCA, which was verified to achieve a near-optimal performance close to the performance upper bound obtained via SDR.
The numerical results indicated that a significant performance gain can be achieved by the NOMA inspired ISAC system than the conventional ISAC system.
Furthermore, compared with the ISAC system with the ideal sensing interference cancellation, the NOMA 
inspired ISAC system can achieve a comparable performance, which provides an efficient scheme in practice.

\bibliographystyle{IEEEtran}
\bibliography{mybib}

\begin{thebibliography}{10}
\providecommand{\url}[1]{#1}
\csname url@samestyle\endcsname
\providecommand{\newblock}{\relax}
\providecommand{\bibinfo}[2]{#2}
\providecommand{\BIBentrySTDinterwordspacing}{\spaceskip=0pt\relax}
\providecommand{\BIBentryALTinterwordstretchfactor}{4}
\providecommand{\BIBentryALTinterwordspacing}{\spaceskip=\fontdimen2\font plus
\BIBentryALTinterwordstretchfactor\fontdimen3\font minus
  \fontdimen4\font\relax}
\providecommand{\BIBforeignlanguage}[2]{{%
\expandafter\ifx\csname l@#1\endcsname\relax
\typeout{** WARNING: IEEEtran.bst: No hyphenation pattern has been}%
\typeout{** loaded for the language `#1'. Using the pattern for}%
\typeout{** the default language instead.}%
\else
\language=\csname l@#1\endcsname
\fi
#2}}
\providecommand{\BIBdecl}{\relax}
\BIBdecl

\bibitem{letaief2019roadmap}
K.~B. Letaief, W.~Chen, Y.~Shi, J.~Zhang, and Y.-J.~A. Zhang, ``The roadmap to
  6{G}: {AI} empowered wireless networks,'' \emph{{IEEE} Commun. Mag.},
  vol.~57, no.~8, pp. 84--90, Aug. 2019.

\bibitem{saad2019vision}
W.~Saad, M.~Bennis, and M.~Chen, ``A vision of 6{G} wireless systems:
  Applications, trends, technologies, and open research problems,''
  \emph{{IEEE} Netw.}, vol.~34, no.~3, pp. 134--142, May 2019.

\bibitem{liu2018toward}
F.~Liu, L.~Zhou, C.~Masouros, A.~Li, W.~Luo, and A.~Petropulu, ``Toward
  dual-functional radar-communication systems: Optimal waveform design,''
  \emph{{IEEE} Trans. Signal Process.}, vol.~66, no.~16, pp. 4264--4279, Aug.
  2018.

\bibitem{liu2018mu}
F.~Liu, C.~Masouros, A.~Li, H.~Sun, and L.~Hanzo, ``{MU-MIMO} communications
  with {MIMO} radar: From co-existence to joint transmission,'' \emph{{IEEE}
  Trans. Wireless Commun.}, vol.~17, no.~4, pp. 2755--2770, Apr. 2018.

\bibitem{liu2020beamforming}
X.~Liu, T.~Huang, N.~Shlezinger, Y.~Liu, J.~Zhou, and Y.~C. Eldar, ``Joint
  transmit beamforming for multiuser {MIMO} communications and {MIMO} radar,''
  \emph{{IEEE} Trans. Signal Process.}, vol.~68, pp. 3929--3944, Jun. 2020.

\bibitem{hua2021optimal}
H.~Hua, J.~Xu, and T.~X. Han, ``Optimal transmit beamforming for integrated
  sensing and communication,'' \emph{arXiv preprint arXiv:2104.11871}, 2021.

\bibitem{liu2021integrated}
F.~Liu, Y.~Cui, C.~Masouros, J.~Xu, T.~X. Han, Y.~C. Eldar, and S.~Buzzi,
  ``Integrated sensing and communications: Towards dual-functional wireless
  networks for 6{G} and beyond,'' \emph{arXiv preprint arXiv:2108.07165}, 2021.

\bibitem{sturm2009ofdm}
C.~Sturm, T.~Zwick, and W.~Wiesbeck, ``An {OFDM} system concept for joint radar
  and communications operations,'' in \emph{Proc. VTC Spring - IEEE 69th Veh.
  Technol. Conf.}\hskip 1em plus 0.5em minus 0.4em\relax IEEE, Apr. 2009, pp.
  1--5.

\bibitem{sturm2011waveform}
C.~Sturm and W.~Wiesbeck, ``Waveform design and signal processing aspects for
  fusion of wireless communications and radar sensing,'' \emph{Proc. {IEEE}},
  vol.~99, no.~7, pp. 1236--1259, Jul. 2011.

\bibitem{tse2005fundamentals}
D.~Tse and P.~Viswanath, \emph{Fundamentals of wireless communication}.\hskip
  1em plus 0.5em minus 0.4em\relax Cambridge university press, 2005.

\bibitem{li2007mimo}
J.~Li and P.~Stoica, ``{MIMO} radar with colocated antennas,'' \emph{{IEEE}
  Signal Process. Mag.}, vol.~24, no.~5, pp. 106--114, Sep. 2007.

\bibitem{liu2017non}
Y.~Liu, Z.~Qin, M.~Elkashlan, Z.~Ding, A.~Nallanathan, and L.~Hanzo,
  ``Non-orthogonal multiple access for 5{G} and beyond,'' \emph{Proc. {IEEE}},
  vol. 105, no.~12, pp. 2347--2381, Dec. 2017.

\bibitem{liu2018multiple}
Y.~Liu, H.~Xing, C.~Pan, A.~Nallanathan, M.~Elkashlan, and L.~Hanzo,
  ``Multiple-antenna-assisted non-orthogonal multiple access,'' \emph{{IEEE}
  Wireless Commun.}, vol.~25, no.~2, pp. 17--23, Apr. 2018.

\bibitem{dinh2010local}
Q.~T. Dinh and M.~Diehl, ``Local convergence of sequential convex programming
  for nonconvex optimization,'' in \emph{Recent Advances in Optimization and
  its Applications in Engineering}.\hskip 1em plus 0.5em minus 0.4em\relax
  Springer, 2010, pp. 93--102.

\bibitem{luo2010semidefinite}
Z.-Q. Luo, W.-K. Ma, A.~M.-C. So, Y.~Ye, and S.~Zhang, ``Semidefinite
  relaxation of quadratic optimization problems,'' \emph{{IEEE} Signal Process.
  Mag.}, vol.~27, no.~3, pp. 20--34, May 2010.

\bibitem{stoica2007probing}
P.~Stoica, J.~Li, and Y.~Xie, ``On probing signal design for {MIMO} radar,''
  \emph{{IEEE} Trans. Signal Process.}, vol.~55, no.~8, pp. 4151--4161, Aug.
  2007.

\bibitem{cvx}
M.~Grant and S.~Boyd, ``{CVX}: Matlab software for disciplined convex
  programming, version 2.1,'' \url{http://cvxr.com/cvx}, Mar. 2014.

\bibitem{mu2021noma}
X.~Mu, Y.~Liu, L.~Guo, J.~Lin, and L.~Hanzo, ``{NOMA}-aided joint radar and
  multicast-unicast communication systems,'' \emph{arXiv preprint
  arXiv:2110.02372}, 2021.

\end{thebibliography}

\end{document}